\documentclass[journal=ancac3,manuscript=article]{achemso}
\usepackage{amsmath}
\usepackage{amsfonts}
\usepackage{graphicx}

\author{Oleksiy Roslyak} 
\email{avroslyak@gmail.com}
\author{Godfrey Gumbs}
\affiliation{Department of Physics and Astronomy,
Hunter College at the City University of New York,
695 Park Avenue New York, NY 10065, USA}
\author{Danhong Huang}
\affiliation{Air Force Research Laboratory, Space Vehicles
Directorate, 
Kirtland Air Force Base, NM 87117, USA}

\date{\today}

\title{Plasma excitations of dressed Dirac electrons in graphene layers}

\begin{document}

\begin{abstract}
Collective plasma excitations of optically dressed
Dirac electrons in single and double graphene layers are calculated
in the RPA approximation.
The presence of circularly polarized light gives rise to an energy gap
$E_g$ between the conduction and valence energy bands. 
Its value may be adjusted by varying the frequency and intensity of the
light, and may reach  values of the gap reported for epitaxially grown graphene and far exceeding that 
caused by spin-orbit coupling.
We report plasmon dispersion relations for various energy gaps and separation
between graphene layers. For a single graphene sheet, we find that plasmon modes
may be excited for larger wave vector and frequency when  subjected to light.
For double layers, we obtained an optical and phonon-like mode
and found that the optical mode is not as sensitive as the phonon-like
mode in the long wavelength limit when the layer separation is varied, for a chosen $E_g$.
The dressed electron plasma although massive still has Dirac origin
giving rise to anomalous plasmon behavior upon crossing the $\omega = \hbar v_F q$ cone.
\end{abstract}


\section{I. Introduction}
\label{sec1}

Recently, several groups have addressed the effects which a circularly polarized
electromagnetic field (CPEF)\,\cite{kibis},
spin-orbit interaction (SOI)\,\cite{SOI} in suspended graphene or the sub-lattice symmetry breaking (SSB)
by an underlying polar substrate\,\cite{SOI-2,li_2009,giovannetti_2007} in epitaxial
graphene may have on the energy band structure of
a graphene sheet. In all cases, a gap is opened between the valence and
conduction bands as well as between the intra-band and inter-band electron-hole
excitation continuum of the otherwise semi-metal Dirac system\,\cite{wunsch,shung1,shung2,Lin}.
As a matter of fact, the interplay between the single-particle excitations in the long wavelength
limit results in dielectric screening of the Coulomb interaction which produces
an undamped plasmon mode that appears in the gap separating the two
types of electron-hole modes forming a continuum. By this we mean that a
self-sustained collective plasma mode is not supported by exciting \emph{either}
the intra-band or inter-band single-particle modes only. Furthermore,
although the Dirac electrons near the $\mathbf{K}$-points now acquire a non-zero effective
mass, one still cannot produce a long wavelength plasmon mode as in the two-dimensional
electron gas (2DEG) by intra-band excitations for intrinsic graphene.

\par
Plasmon excitations in semi-metallic graphene were studied extensively theoretically in single and
multilayer graphene. However, their experimental observations are limited so far to either
inelastic scattering of highly energetic particles (faster than $10^5 \; m/s$) or
indirect observation through plasmoritons (a quasiparticle formed by resonant
coupling of renormalized hole states with the plasmons).
Plasmons in graphene gapped by SOI or by a polar substrate have been investigated by
Wang and Chakraborty\,\cite{SOI} as well as Pyatkovskiy\,\cite{SOI-2}.
Because the SOI is small ($\sim 0.05$ meV), its effect on dielectric screening of the
Coulomb interaction requires either high resolution spectroscopy or low electron density
in the conduction band to be observed.
The effect of SOI on the plasmon excitations
in a 2DEG were also studied by several authors\,\cite{gumbs1,manvir,gumbs2}.
On the other hand, the gap between the valence and
conduction bands near the Fermi level when circularly polarized light is
applied to graphene could be manipulated, unlike the intrinsic gap due to
SOI, depending on the intensity and frequency of the light. For a CO$_2$ laser
with a  70 W output power and wavelength of 10 $\mu$, the energy gap is about
100 meV, compared with 0.08 meV induced by SOI\,\cite{SOI}.
The exact gap opening mechanism in epitaxial graphene is still under debate\,\cite{rotenberg_2008}.
The problem is rather technical and arises from the fact that the graphene
sample  sits on top of a buffer layer which provides an additional mid gap
level, thus obscuring the exact energy dispersion curve and
 requires numerical  \emph{ab initio} calculations\,\cite{kim_2008}.
This ambiguity stimulated  discussion regarding the symmetry breaking
gap\,\cite{zhou_2007} versus the effects due to  electron-electron
interaction\,\cite{bostwik_2007}. This also not only indicates the importance
of theoretical investigation of   plasmon in gapped graphene,
but also an alternative means of providing the gap.
Since SOI is too weak to observe its effect, dressing of the Dirac electrons with
CPEF is seems to be an alternative worthy of further  investigation.

\par
\begin{figure}
\centering
\includegraphics[width=0.8\columnwidth]{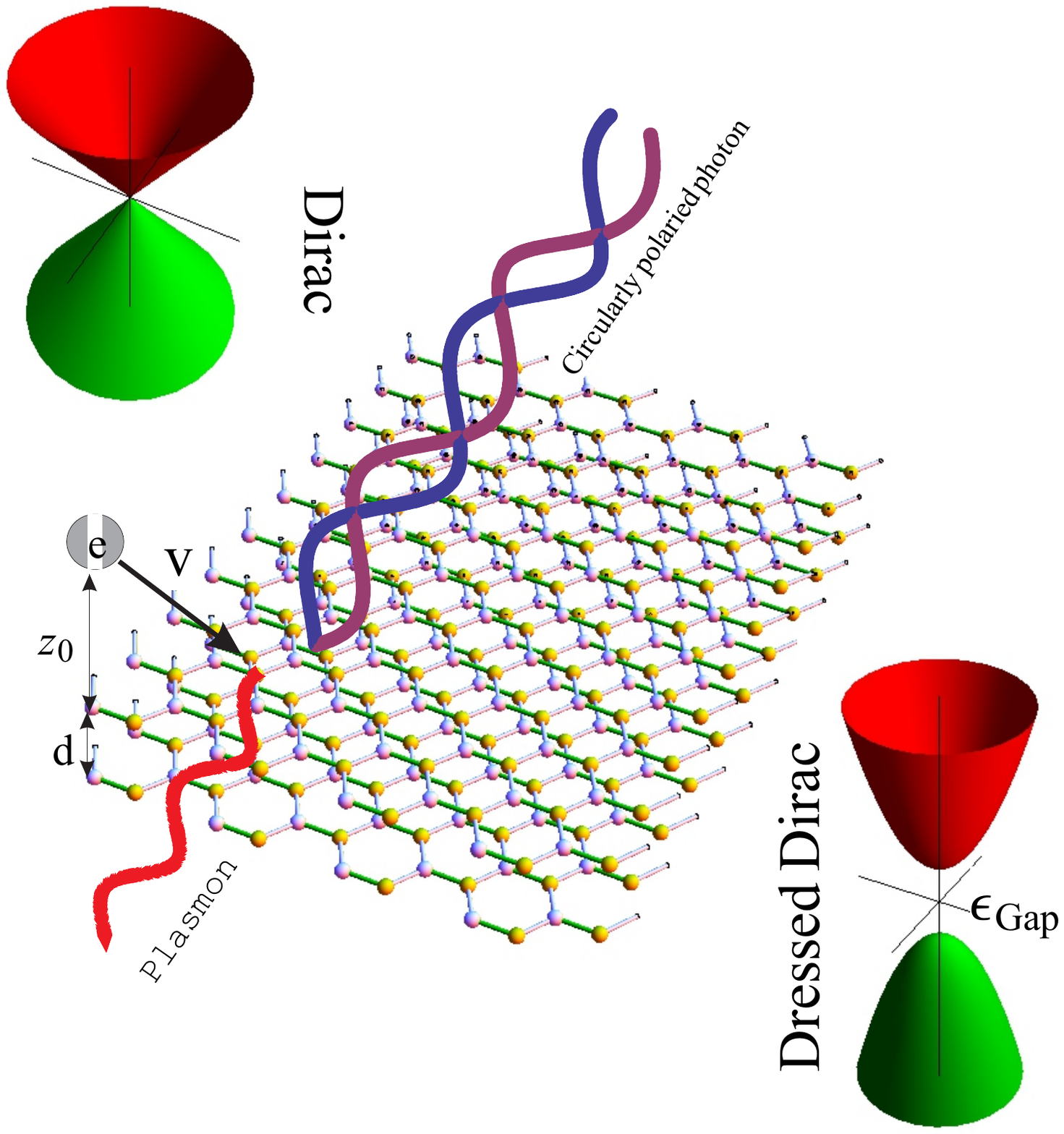}
\caption{ (Color online)
Schematic illustration of the pump-probe consisting of circularly
polarized light and a beam of energetic charged particles. On the left is the
energy band structure for conventional Dirac fermions near the K-point. On the
right is the corresponding band structure for dressed Dirac fermions.}
\label{FIG:0}
\end{figure}

The value of this dynamic energy gap is comparable with that   induced by a polar substrate, but contrary to the
latter, it  is highly adjustable. As a matter of fact, those two gap-opening mechanisms
may even work conjunctively. In that case, the circular polarization of the electromagnetic
field may  no longer be required provided that it is tuned to be in resonance with the SSB
gap. A possible experiment to measure the plasmon frequencies in single and double layer graphene
could employ electron energy loss spectroscopy (EELS)\,\cite{p6.1,ggumbs,antonio}, as shown
schematically in Fig.\,\ref{FIG:0}. Essentially,  this would be a pump-probe,
with the electron beam being the probe effected by the polarized light whose role
is to open the gap.  The Raman shift of the scattered electrons provides both particle-hole
and plasmon excitation frequencies, which are usually characterized by their spectral weight,
a quantity that  depends on the
transferred energy $\hbar \omega$ and momentum $\hbar q$.

\par
The remainder of this paper  is organized as follows. In Sec. II, we present
the model Hamiltonian for  dressed Dirac electrons by a CPEF as well as
corresponding energy eigenvalues  and wave-functions. These results  are then used
for calculating  the   polarization function and plasmon dispersion relation in the RPA.
Our numerical results  are presented in Sec. III but, for convenience, we
briefly summarize  the highlights of our calculations. For a single graphene layer, in
the presence of CPEF, we calculate analytically the real and imaginary
parts of the noninteracting polarization function , whose imaginary part provides the
particle-hole spectral weight. As a piecewise continuous function, it is expressed
in terms of region functions which provide a convenient way for determining the  Landau
damping of the plasmon excitations. By  increasing the  intensity of CPEF, the Dirac
electrons become more dressed by  photons, providing them with a larger and larger
effective mass. Regions in $\omega-q$ space, which are non-existent for conventional
Dirac electrons, emerge thereby making the plasmon behavior to resemble that of conventional 2DEG.
However, the resemblance is limited, and we show there is no complete crossover to 2DEG behavior.
The behavior of the plasmons is definitely determined by the presence of the Dirac cone
$\omega = v_F q$, where $v_F$ is the Fermi velocity.  This indicates that although the
dressed electron plasma  is massive, it  still has Dirac origin. We also investigate
graphene double layers subjected to a CPEF with two plasmon branches given as resonances
of the real part of the inverse dielectric function. Two cases are investigated: the symmetric
configuration when   both layers are gapped and the asymmetric case when the gap is induced on one
of the layers. The latter case may be obtained by compensating CPEF with the SSB on the given layer.
The main derived effect on the acoustic plasmon branch is that this mode shows a
gain for the symmetric case, but a loss for the asymmetric case. This applies for a
sufficiently intense laser field,  over a restricted range of frequency and
wavelength\,\cite{antonio}. In Sec. IV, we offer
some remarks concerning experimental observation of the plasma excitations.

\section{II.\label{sec2} Model and Formulation of the Problem}

Let us consider several graphene layers separated by distance $d$.
One of the layers may be epitaxially grown on a SiC substrate.
The structure is optically pumped by an idealized single mode CPEF
characterized by an associated energy $N_0 \hbar \omega_0$
where $N_0$ is the average number of photons in the optical field.
It is assumed to be a function of the external pump intensity.
We   also take it that all graphene layers are in the node of the
optical field and we can neglect retardation effects. This is valid
provided that $d \ll 2 \pi c/\omega_0$. The objective is to find the plasmon
dispersion branches $\omega_{\pm}(q)$ for double layer graphene. Those are
given by the poles of the inverse dielectric function. A single layer may
be described by the same formalism by setting the inter-layer distance to infinity.
By setting the origin $z=0$ at one of the layers,
we may follow the procedure presented in Ref.\cite{EELS_SSC} to obtain the inverse dielectric function as:

\begin{equation}
\epsilon^{-1}(z,z^\prime;q,\omega)= \delta(z-z^\prime)   +  \sum_{j,j^\prime=0,1}
\delta\left({z-(j-1)d}\right)  v_c(jd,j^\prime d)\Pi_{jj^\prime}^{(0)}(q,\omega)
\end{equation}
where $\delta(z-z^\prime)$ is the Dirac delta function and polarization matrix is

\begin{equation}
\left({
\begin{array}{cc}
\Pi_{11} & \Pi_{12}\\
\Pi_{21} & \Pi_{22}
\end{array}
}\right)
=
\frac{1}{\epsilon(q,\omega)}
\left({
\begin{array}{cc}
\Pi^{(0)}_{11} \left({1-v_c(q)\Pi^{(0)}_{22} }\right) & v_c(q)\ e^{-qd}\Pi^{(0)}_{11} \Pi^{(0)}_{22}\\
v_c(q)\ e^{-qd}\Pi^{(0)}_{11} \Pi^{(0)}_{22} & \Pi^{(0)}_{22} \left({1-v_c(q)\Pi^{(0)}_{11}}\right)
\end{array}
}\right)
\end{equation}
It is expressed in terms of the generalized dielectric function

\begin{equation}
\label{EQ:DETERMINANTSINGLE}
\epsilon(q,\omega)= 1-v_c(q)\left\{\left[\Pi^{(0)}_{11} +\Pi^{(0)}_{22}  \right]+\left[
\left( 1 - e^{-2qd} \right)v_c(q)\Pi^{(0)}_{11} \Pi^{(0)}_{22}  \right] \right \} \ ,
\end{equation}
where $v_c(q) = 2 \pi e^2/\epsilon_s q$ with $\epsilon_s\equiv 4\pi\varepsilon_0\epsilon_b$,
$\epsilon_b$ being the average background dielectric constant.
The plasmon resonances correspond to the poles of the inverse dielectric function which are
given by the solutions of $\epsilon(q,\omega)= 0$. Additionally,
the plasmon frequency incorporates its a  relaxation  rate so that $\omega\to \omega + i \eta$.
Also, $\Pi^{(0)}_{jj}(q,\omega+ i \eta)$ is the noninteracting polarization function for
the $j^{\text{th}}$ graphene layer.
The inter-layer polarization function $\Pi^{(0)}_{12}$ is not included in our formalism
due to negligible interlayer tunneling.

\par
The key ingredient of the above equations is the polarization function. This requires specifics
from the model given by the Hamiltonian at the two inequivalent $\mathbf{K},\mathbf{K}^\prime$ points
in the graphene dispersion:
\begin{gather}
\label{EQ:HAMILTONIAN}
H=\mathcal{H}_{JC}+\mathcal{H}_D\\
\label{EQ:JAMESCUMMINSHAMILTONIAN}
\mathcal{H}_{JC} = \hbar \omega_0 a^\dag a + \frac{\Delta \mu_{A,B}}{2} \sigma_3
-\frac{W_0}{2 \sqrt{N_0}} \left({\sigma_{+} a + \sigma_{-} a^\dag}\right)\\
\label{EQ:DIRACHAMILTONIAN}
\mathcal{H}_D = \hbar \it{v}_F \left({\left({\sigma_{+} + \sigma_{-}}\right) k_x \pm i\left({\sigma_{-} - \sigma_{+}}\right) k_y}\right)\ .
\end{gather}
Here, the Jaynes-Cummings Hamiltonian \eqref{EQ:JAMESCUMMINSHAMILTONIAN} is described by the interaction
third term governed by the amplitude of the electric filed $E_0 = \sqrt{N_0 \hbar \omega_0/\epsilon_0V}$,
and $V$ is the quantization volume of the CPEF. The energy of an electron in rotational motion
induced by the field is denoted as $W_0 = 2 v_F e E_0/\omega_0 \sim \sqrt{N_0}$ and assumed to be much
smaller than the energy of the optical filed itself, i.e., $W_0/\hbar \omega_0 \ll 1$. The
second term in \ref{EQ:JAMESCUMMINSHAMILTONIAN} takes account of the breaking of
the inversion symmetry between $A$ and $B$ sub-lattices. Also, $\Delta \mu_{A,B}$
stands for the difference in on-site energies for the sub-lattices. It may be induced if
the graphene is grown epitaxially on a substrate, and may acquire values as large as
$260 \; meV$ ($4775 \; nm$), see, for instance, Zhou et al.\cite{zhou_2007} and references therein.
On the smaller scale ($\sim 10^{-1} \; meV$) the symmetry breaking term accounts for spin-orbit interaction.

\par
The Dirac part of the Hamiltonian is given by \ref{EQ:DIRACHAMILTONIAN}, with $\pm$ standing for
$\mathbf{K},\mathbf{K}^\prime$ points correspondingly. The operators involved in the above Hamiltonian
are acting in the joined electro/photon space via the following relations:

\begin{eqnarray*}
\sigma_{+} = \vert \uparrow \rangle \langle \downarrow \vert \; ; &
\sigma_{-} = \vert \downarrow \rangle \langle \uparrow \vert \; ; \\
\sigma_{+} \vert \uparrow, N_{0} \rangle = 0 \; ; & \sigma_{+} \vert \downarrow, N_{0} \rangle = \vert \uparrow, N_{0} \rangle \\
\sigma_{-} \vert \downarrow, N_{0} \rangle = 0\; ; & \sigma_{-} \vert \uparrow, N_{0} \rangle = \vert \downarrow, N_{0} \rangle \\
a \vert \downarrow \uparrow, N_{0} \rangle = \sqrt{N_{0}} \vert \downarrow \uparrow, N_{0}-1 \rangle\;
; & a^\dag \vert \downarrow \uparrow, N_{0} \rangle = \sqrt{N_{0}+1} \vert \downarrow \uparrow, N_{0}+1 \rangle
\end{eqnarray*}
with $\vert{\uparrow \downarrow}\rangle$ denoting the Dirac pseudo-spin basis.
The Jaynes-Cummings part can be readily diagonalized by switching to the dressed electron basis

\begin{gather}
\label{EQ:DRESSEDBASIS}
\left({
\begin{array}{c}
\vert{+,N_{0}}\rangle\\
\vert{-,N_{0}}\rangle
\end{array}
}\right)=
\left({
\begin{array}{cc}
\cos \left({\Phi_{N_{0}/2}}\right) & \sin \left({\Phi_{N_{0}}/2}\right)\\
-\sin \left({\Phi_{N_{0}}/2}\right) & \cos \left({\Phi_{N_{0}}/2}\right)
\end{array}
}\right)
\left({
\begin{array}{c}
\vert{\uparrow,N_{0}}\rangle\\
\vert{\downarrow,N_{0}+1}\rangle
\end{array}
}\right)\\
\notag
\cos{\left({\Phi_{N_{0}}/2}\right)} = \sqrt{\frac{\Omega_{N_{0}} + \Delta}{2 \Omega_{N_{0}}}}\\
\notag
\sin{\left({\Phi_{N_{0}}/2}\right)} = \sqrt{\frac{\Omega_{N_{0}} - \Delta}{2 \Omega_{N_{0}}}}
\end{gather}
where $\Omega_{N_{0}}^2 = \Delta^2 + 4 W^2_0 (N_{0}+1)/N_{0}$ and the detuning between the
microcavity and the symmetry breaking induced gap is denoted by
$\Delta = \hbar \omega - \Delta \mu_{A,B}$. In the new basis,  the Hamiltonian \ref{EQ:HAMILTONIAN}
within the  small dressing regime $W_0 \ll \Delta$ assumes the following form

\begin{equation}
\label{EQ:DRESSEDHAMILTONIAN}
H = \mathbb{I} N_0 \hbar \omega_0 + \frac{E_g}{2} \sigma_3 +  \mathcal{H}_D
\end{equation}
where $E_g = \sqrt{W^2_0 + \Delta^2}-\hbar \omega_0$ is the energy gap between valence
and conduction bands. The gap governs the nature of
dressed Dirac electrons metal-insulator transition\cite{kibis}.
One may obtain the solution of the above equation in a  straightforward way, giving

\begin{gather}
\label{EQ:EIGENVALUES}
E_{\pm,\mathbf{k}} = N_0 \hbar \omega_0 \pm \sqrt{(E_g/2)^2 + (\hbar v_F k)^2}\\
\label{EQ:EIGENFUNCTIONS}
\left({
\begin{array}{c}
\vert{+,{\mathbf{k}} }\rangle\\
\vert{-,{\mathbf{k}} }\rangle
\end{array}
}\right) = \frac{\texttt{e}^{i \mathbf{kr}}}{\sqrt{1 + \alpha^2_{N_{0},k}}}
\left({
\begin{array}{cc}
1 & \mathcal{A}_{N_{0},k}\texttt{e}^{i \theta_{\mathbf{k}}}\\
\mathcal{A}_{N_{0},k} & -\texttt{e}^{i \theta_{\mathbf{k}}}
\end{array}
}\right)
\left({
\begin{array}{c}
\vert{+\frac{1}{2},N_0}\rangle\\
\vert{-\frac{1}{2},N_0}\rangle
\end{array}
}\right) \\
\mathcal{A}_{N_{0},k} = \frac{\hbar v_F k}{\sqrt{(\hbar v_F k)^2+(E_g/2)^2}+E_g/2}
\end{gather}

\begin{figure}
\centering
\includegraphics[width=1.0\columnwidth]{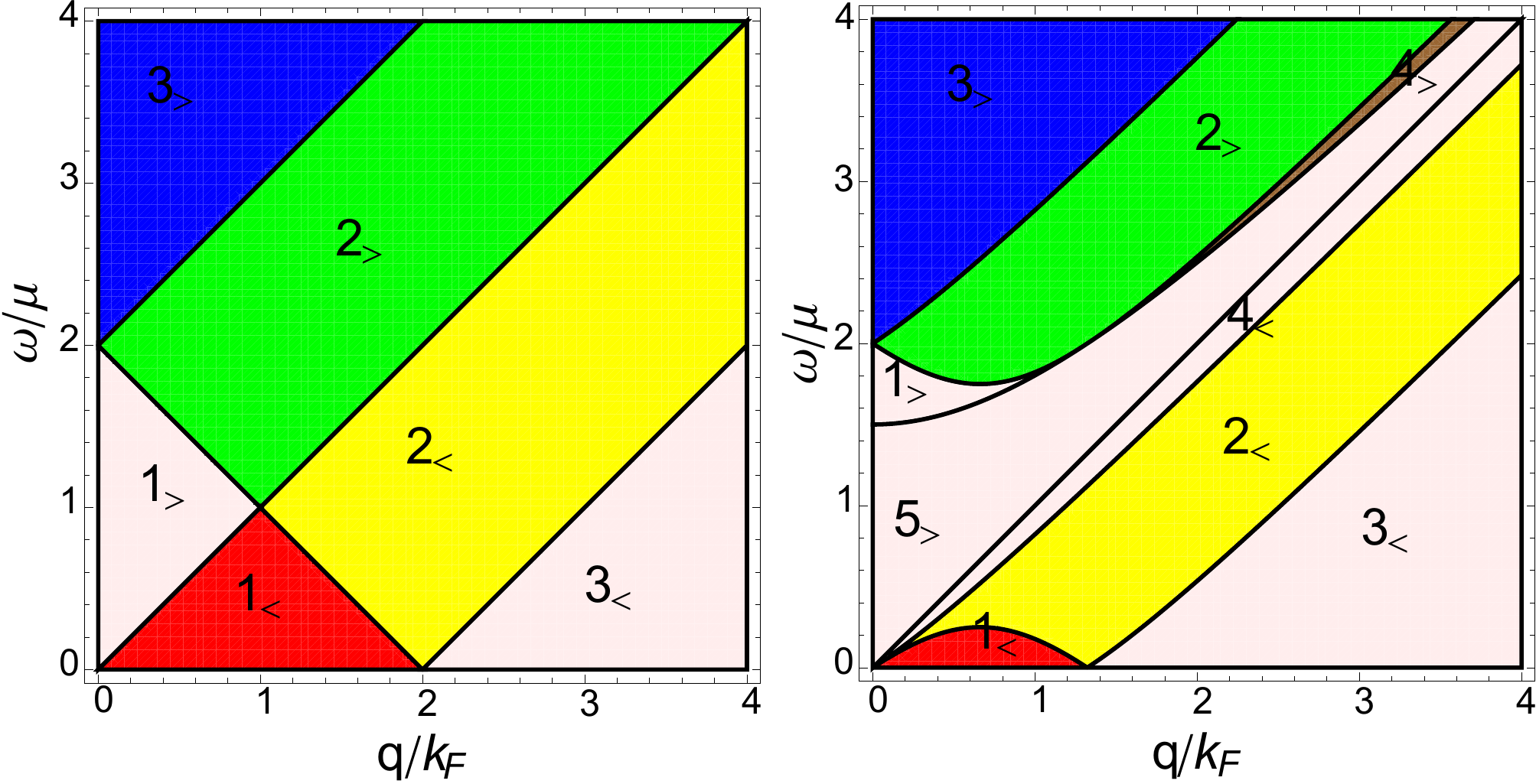}
\caption{ (Color online)
Labeled regions for analytic  results for the polarization function in the  case of
conventional graphene (zero energy gap, left panel) and for the  case when there is an
induced gap   $E_g/\mu = 1.5$ (right panel).}
\label{FIG:6}
\end{figure}

It is now a simple matter  to show that conventional Dirac fermions correspond to
the off-resonance limit of the dressed states. Given the dressed states, the noninteracting
polarization assumes the Lindhard form

\begin{gather}
\label{EQ:NETPOLARIZATION}
\Pi^{(0)}_{jj}(q,\omega) = \Pi^{(0),1}_{jj}(q,\omega)+\Pi^{(0),2}_{jj}(q,\omega)\\
\begin{array}{cc}
\Pi^{(0),1}_{jj}(q,\omega) = -\frac{1}{\pi}
\sum \limits_{\alpha = \pm} \int \limits_{0}^{\infty}
k dk \mathcal{I}^{\alpha -}(k,q,\omega) ;
&
\Pi^{(0),2}_{jj}(q,\omega) = \frac{1}{\pi}
\sum \limits_{\alpha,\beta = \pm} \int \limits_{0}^{\Re e\ \sqrt{\mu^2 - (E_g/2)^2}/\hbar v_F}
k dk \mathcal{I}^{\alpha \beta}(k,q,\omega)\\
\notag
\mathcal{I}^{\alpha \beta}(k,q,\omega) = \int \limits_{0}^{2 \pi} d \phi
\frac{\mathcal{F}^{\beta}_{\mathbf{k},\mathbf{k+q}}}
{ \hbar \omega+\alpha E_{\mathbf{k}}-\alpha \beta E_{\mathbf{k+q}}};
&
\mathcal{F}^{\beta}_{\mathbf{k},\mathbf{k+q}} =
\frac{1}{2} \left({
1+ \beta \frac{\hbar^2 v^2_F \mathbf{k} \cdot \left({\mathbf{k+q}}\right) + (E_g / 2)^2}
{E_{\mathbf{k}} E_{\mathbf{k+q}}}
}\right)
\end{array}
\end{gather}
The two terms in \ref{EQ:NETPOLARIZATION} stand for the inter- and intra-band polarizations,
respectively. Also,  $\mu$ is the chemical potential whose zero value is in the middle of the gap ($E_g$).
In calculating the polarization, we assumed that  for chosen $N_0$, each point on the dispersion
curve $E_{\mathbf{k}}$ might be occupied by no more than one electron. This assumption is valid for
small energy gap. Otherwise, the quasi-equilibrium density matrix acquires non vanishing off-diagonal
elements thus substantially complicating the statistical properties of the dressed states.
The noninteracting polarization \ref{EQ:NETPOLARIZATION} in   gaped graphene has already been
calculated on both the real and imaginary frequency axes\cite{quaimzadeh_2009,SOI-2}
which essentially yield the same result. Since it plays a crucial role in this paper, we
give its full form along the real frequency axis as

\begin{gather}
\label{EQ:POLARIZATION}
\Pi^{(0)}_{j}(q,\omega) = -\frac{2 \mu}{\pi \hbar^2 v^2_F} +
\frac{q^2}{4 \pi \sqrt{\vert{\hbar^2 v_F^2 q^2 - \hbar^2 \omega^2}\vert}}
\times [\\
\notag
\left({i G_>(x_{1,-})- i G_>(x_{1,+}))}\right)1_< +
\left({ G_<(x_{1,-}) + i G_>(x_{1,+}))}\right)2_< +\\
\notag
\left({G_<(x_{1,+}) + G_<(x_{1,-}))}\right)3_< +
\left({G_<(x_{1,-}) - G_<(x_{1,+}))}\right)4_< +
\left({G_>(x_{1,+}) - G_>(x_{1,-}))}\right)1_> +\\
\notag
\left({G_>(x_{1,+}) + i  G_<(x_{1,-}))}\right)2_> +
\left({G_>(x_{1,+}) - G_>(-x_{1,-}) - i \pi (2-x_0^2))}\right)3_> +\\
\notag
\left({G_>(-x_{1,-}) + G_>(x_{1,+}) - i \pi (2-x_0^2))}\right)4_> +
\left({G_0(x_{1,+}) - G_0(x_{1,-}) )}\right)5_> ]
\end{gather}
Here, the following notations for the region functions have been introduced:

\begin{eqnarray*}
x_0&=&\sqrt{1+\frac{E_g^2}{\hbar^2 v_F^2 q^2-\hbar^2\omega^2}},\nonumber\\
x_{1,\pm}&=&\frac{2 \mu \pm\hbar \omega}{\hbar v_F q},\nonumber\\
x_{2,\pm}&=&\sqrt{\hbar^2 v_F^2 (q\pm k_F)^2 + (E_g/2)^2},\nonumber\\
x_{3}&=&\sqrt{\hbar^2 v^2_F q^2 + E^2_g},\nonumber\\
G_<(x)&=&x\sqrt{x_0^2-x^2}-(2-x_0^2)\cos^{-1}(x/x_0),\nonumber\\
G_>(x)&=&x\sqrt{x^2-x_0^2}-(2-x_0^2)\cosh^{-1}(x/x_0),\nonumber\\
G_0(x)&=&x\sqrt{x^2-x_0^2}-(2-x_0^2)\sinh^{-1}(x/\sqrt{-x_0^2}),
\end{eqnarray*}
and the regions are shown in Fig.\ \ref{FIG:6} and are defined as

\begin{eqnarray*}
1_< = &\theta \left({ \mu - x_{2,-} - \hbar \omega }\right)\\
2_< = &\theta \left({ - \hbar \omega - \mu + x_{2,-} }\right)
\theta \left({ \hbar \omega + \mu - x_{2,-} }\right)  \theta \left({ \mu + x_{2,+} - \hbar \omega }\right)\\
3_< = &\theta \left({ -\mu + x_{2,-} - \hbar \omega }\right)\\
4_< = &\theta \left({ \hbar \omega + \mu - x_{2,+} }\right)  \theta \left({\hbar v_F q - \hbar \omega }\right)  \\
1_> =  &\theta \left({ 2 k_F - q }\right)  \theta \left({ \hbar \omega - x_3 }\right)
\theta \left({\mu + x_{2,-} - \hbar \omega }\right) \\
2_> =  &\theta \left({ \hbar \omega - \mu - x_{2,-} }\right)
\theta \left({ \mu + x_{2,+} - \hbar \omega }\right) \\
3_> =  &\theta \left({ \hbar \omega - \mu - x_{2,+} }\right)\\
4_> =  &\theta \left({q - 2 k_F }\right)  \theta \left({ \hbar \omega - x_3 }\right)
\theta \left({ \mu +x_{2,-} - \hbar \omega }\right) \\
5_> =  &\theta \left({ \hbar \omega - \hbar v_F q}\right)  \theta \left({x_3 - \hbar \omega}\right)
\end{eqnarray*}

For the purposes of optical spectroscopy, it is advisable to study the long wavelength limit
by expanding the polarization function in \ref{EQ:POLARIZATION} in powers of $q$.
For small $q/k_F$, the undamped plasmons exist in the spectral regions $1_>$ and $5_>$.
By expanding the polarization up to the fourth order in $q$, we obtain

\begin{equation}
\label{EQ:POLARIZATIONREDUCED}
\Pi^{(0)}_{jj}(q,\omega) = \frac{q^2}{4 \hbar^3 \pi \omega^3} \times
\left\{{
\begin{array}{cc}
4 \hbar \mu \omega + \left({E_g^2 + \hbar^2 \omega^2}\right)
\log \left[{\frac{2 \mu - \hbar \omega}{2 \mu  + \hbar \omega}}\right]  + O(\frac{q^4}{\omega^4}); &
1_>,5_>,4_<;
\\
 -\frac{16 \hbar \mu \omega^3}{v^2_F q^2} - 4 \hbar \mu \omega - \left({E_g^2 + \hbar^2 \omega^2}\right)
\log \left[{\frac{2 \mu - \hbar \omega}{2 \mu  + \hbar \omega}}\right]  + O(\frac{q^4}{\omega^4}); & 1_<,2_<.
\end{array}
}\right.
\end{equation}
For small frequency $\omega$, the following identity applies, 
$ \left({E_g^2 + \hbar^2 \omega^2}\right)
\log \left[{\frac{\mp 2 \mu \pm \hbar \omega}{2 \mu + \hbar \omega}}\right]
\approx - \hbar E_g^2 \omega / \mu$, 
which we employed to  further simplify
the polarization \eqref{EQ:POLARIZATIONREDUCED} as

\begin{gather}
\label{EQ:POLARIZATIONOVERSIMPLIFIED}
\Pi^{(0)}_{jj}(q,\omega ) =
\frac{q^2 \mu }{\pi \hbar^2 \omega^2} 
\left({1-\frac{E_g^2}{4 \mu^2}}\right)
\left({1_> +5_> + 4_<}\right) -\\
\notag
-\left({ 
\frac{4 \mu}{\pi \hbar^2 v^2_F}-
\frac{q^2 \mu }{\pi \hbar^2 \omega^2} 
\left({1-\frac{E_g^2}{4 \mu^2}}\right)
}\right)
\left({1_< + 2_<}\right)
\end{gather}

Note that smallness of $\omega$ is automatically assured in the regions $4_<,1_<,2_<$.
In these regions, the dominant excitations are of the particle-hole type, 
due to nonzero imaginary part of the polarization.
In the long wavelength limit, the plasmon frequency for a  single
graphene layer in the vacuum is obtained by solving for the zeros
of $\epsilon(q,\omega)$ in \ref{EQ:DETERMINANTSINGLE}, 
along with $\Pi^{(0)}_{11} = \Pi^{(0)} (q,\omega)$ given by \ref{EQ:POLARIZATIONOVERSIMPLIFIED}
and  $\Pi^{(0)}_{22}$ set to zero, yielding

\begin{equation}
\label{EQ:PLASMONSINGLELAYER}
\omega^2_p = q \mathcal{P}(E_g,\omega_p) \ ,
\end{equation}
where we introduced the plasmon region function

\begin{gather}
\label{EQ:PLASMONREGIONFUNCTION}
\mathcal{P}(E_g,\omega_p) = \frac{2 \mu e^2}{\hbar^2 \epsilon_0} ({1-\frac{E_g^2}{4 \mu^2}}) \left({ (1_> 
+ 5_> + 4_<)\vert_{\omega \rightarrow \omega_p} + 
\frac{q}{q+80 \pi \mu/\hbar v_F}(1_< + 2_<)\vert_{\omega \rightarrow \omega_p} }\right) \ .
\end{gather}
In the above expression we used $e^2/4 pi \epsilon_0 \hbar v_F = 2.5$
It is also straightforward to show that the region function disappears in the region $1_<$, 
thus supporting Landau damped plasmons in the region $2_<$ only.
We also note that for conventional graphene and the 2DEG, the plasmon region
function \ref{EQ:PLASMONREGIONFUNCTION} is replaced by constants
$2 \mu e^2/\hbar^2 \epsilon_s$
and  $m^\star v_F^2 e^2/\hbar^2 \epsilon_s$, respectively.

\par
For the symmetric double layer configuration, one obtains two
undamped and two Landau damped plasmon branches whose frequencies
in the long wavelength limit obey

\begin{equation}
\omega^2_{\pm} = q \texttt{e}^{- q d}\left({\texttt{e}^{q d} \pm 1}\right) \mathcal{P} (E_g,\omega_{\pm})
\end{equation}
where $\pm$ stands for the bonding/anti-bonding plasmon modes, when the polarizations in the 
two layers work together/against each other in producing the charge density oscillations.
Two independent parameters are $q d$ and $E_g$. That is the interplay between the inter-layer
distance and the plasmon region function determines the plasmon branches.
In the  limit of large inter-layer separation,
$q d \gg 1$, both branches are reduced to \ref{EQ:PLASMONREGIONFUNCTION}.
Conventionally\,\cite{dassarma}, those regimes are referred to as  intermediate
and weak coupling limits.
We avoid  using these terms so as not to confuse with the light-matter coupling.
On the other hand,  for small inter-layer separations,  $q d \ll 1$, we have

\begin{gather}
\label{EQ:DLOPTICAL}
\omega^2_+ = q  \mathcal{P} (E_g,\omega_{\pm}) \left({2- q d}\right)\\
\label{EQ:DLACOUSTICAL}
\omega^2_- = q^2 \mathcal{P} (E_g,\omega_{\pm}) d
\end{gather}
Due to linear dependence of the undamped plasmon frequency on the wave vector $q$,
the anti-bonding mode is often referred to as acoustical branch.
The other undamped plasmon branch  has larger frequency and
is named the optical branch. To the leading order in $q$, this branch
is independent of  inter-layer separation and is completely determined by
the plasmon region function. It is always higher in frequency than the acoustical
branch and usually (but not always) experiences Landau damping in
the  long wavelength limit.

\par
The asymmetric case is also easily obtained in the long wavelength limit, i.e.,

\begin{equation}
\omega^2_{\pm} = \frac{1}{2} q \texttt{e}^{- q d} \left({ \texttt{e}^{q d} (\mathcal{P}_{1,\pm}+\mathcal{P}_{2,\pm}) \pm
\sqrt{\texttt{e}^{2 q d} (\mathcal{P}_{1,\pm} - \mathcal{P}_{2,\pm})^2 + 4 \mathcal{P}_{1,\pm} \mathcal{P}_{2,\pm}} }\right) \ ,
\end{equation}
where we have introduced the short hand notation 
$\mathcal{P}_{1(2),\pm} = \mathcal{P}(E_{g,1(2)},\omega_{\pm}) $,
with $E_{g,1(2)}$ denoting the gap induced on the first (second) graphene layer.
In the small inter-layer separation limit, these eigenfrequencies become

\begin{gather}
\omega^2_{+} = q \left({\mathcal{P}_{1,+}^2 +\mathcal{P}_{2,+}^2- 
2 \mathcal{P}_{1,+}\mathcal{P}_{2,+} (q d -1)}\right)\left({\mathcal{P}_{1,+} + \mathcal{P}_{2,+}}\right)^{-1}\\
\omega^2_{-} = 2 d q^2 \mathcal{P}_{1,-}\mathcal{P}_{2,-} \left({\mathcal{P}_{1,-} + \mathcal{P}_{2,-}}\right)^{-1}\ .
\end{gather}
When one moves away from the long wavelength limit, $\epsilon(q,\omega)=0$  from 
\ref{EQ:DETERMINANTSINGLE} in conjunction  with \eqref{EQ:POLARIZATION} become nonlinear, 
with no closed-form analytic
solutions. 
Consequently,  numerical simulations are needed. 
Our numerical results are the subject of the next section.

\section{III. Numerical results and Discussion}
\label{sec3}

In our numerical calculations, we scaled energies  in units of the chemical
potential $\mu$ and the wave number in units of the Fermi wave number $k_F$.
The frequency of the electromagnetic field along with its intensity determines
the energy gap $E_g$ between the valence and conduction bands. Consequently,
this is the only parameter we use to specify the role played by the external
laser pump field.  All our calculations for the collective plasma
excitations were carried out at $T=0 \; K$.
\par

\begin{figure}
\centering
\includegraphics[width=0.5\columnwidth]{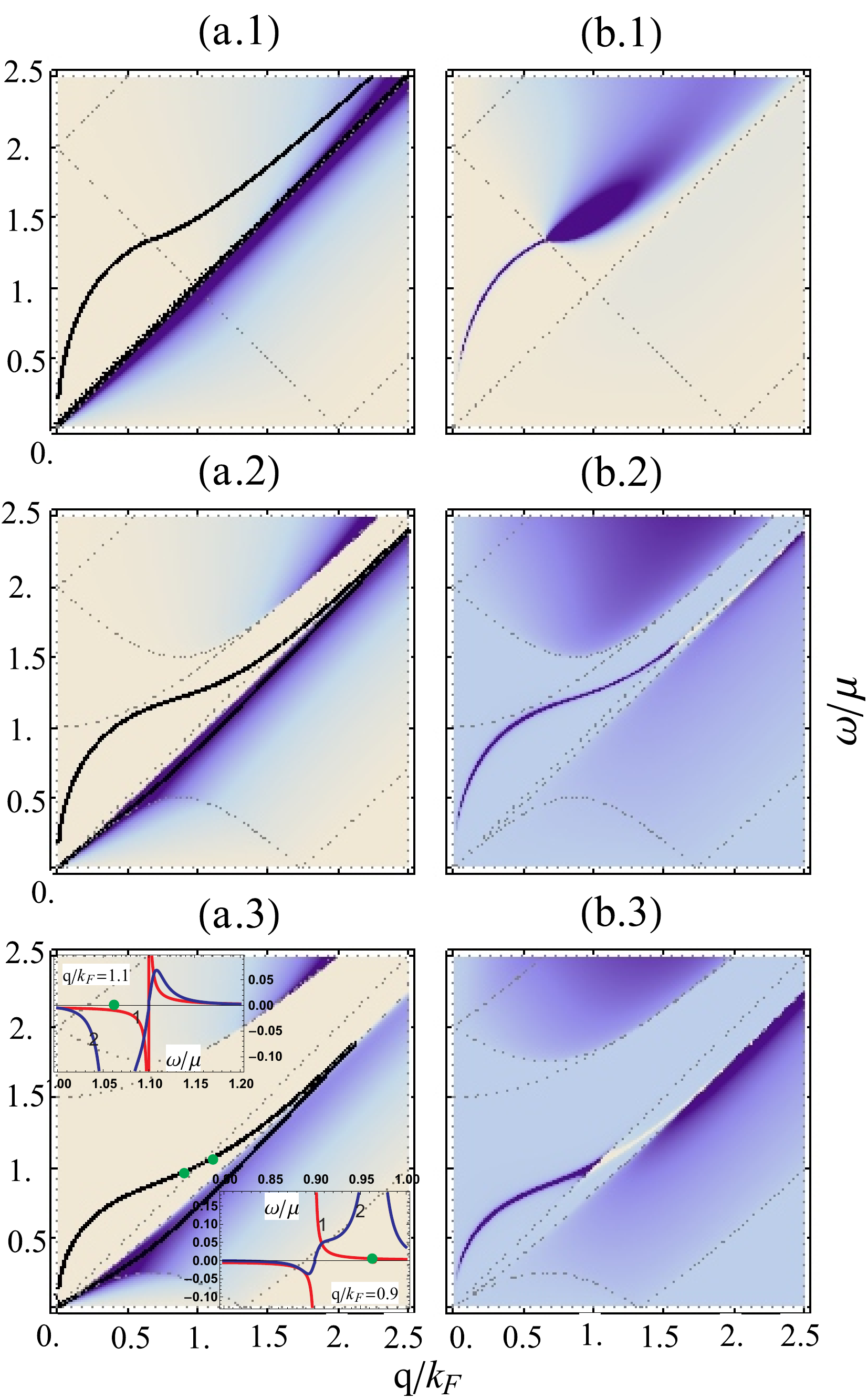}
\caption{ (Color online) The plasma excitations and their damping are presented for
a single graphene layer. In rows (1,\,2,\,3), the energy gaps chosen are $E_g/\mu = \{{0.0,\,1.0,\,1.5}\}$,
respectively. In the panels labeled (a), we plot $\Im m\  \Pi^{(0)}(q,\omega)$
as well as $\Re e\ \epsilon (q,\omega) = 0$ (with $d\to \infty$).
Panels (b) were obtained by plotting the   spectral weight   $-(1/\pi) \Im m\ \Pi (q,\omega - i \gamma_q)$
which involves the imaginary part of the RPA form of the polarization function.
In panels (a),  the darker the shaded regions, the larger is    the value of $\Im m\ \Pi^{(0)}(q,\omega)$,
corresponding to Landau damping.  The black solid curves
in panels (a) show the plasmon dispersion obtained by solving  $\Re e\ \epsilon (q,\omega) = 0$.
The finite lifetime of the plasmon modes represented by the thin solid black curves in panels (b),
was  included in  the polarization function to take account of phonon-like scattering and was determined
by solving   $\Im m\ \Pi (q,\omega + i \gamma_q) = 0$.
The auxiliary dashed lines show $n_>$ and $n_<$ regions.
The upper and lower insets of (a.3) correspond to the fixed values of $q$ so that the plasmon branch is below and
above $\omega = \hbar v_F q$ line. The green dot indicates the position of the plasmon resonance $\omega_p$. Red (1) curve
indicates $\Im \Pi^{(0)}$, while blue (2) curve stands for  $\Im \Pi$.
}
\label{FIG:1}
\end{figure}

We begin this section by considering the  plasmon dispersion for conventional
Dirac electrons in graphene obtained by solving $\Re e\ \epsilon(q,\omega)=0$ with
the use of \ref{EQ:DETERMINANTSINGLE}. 
Our results are  presented in
\ref{FIG:1}(a.1) and  are superimposed onto the particle-hole
modes by plotting $\Im m\ \Pi^{(0)} (q, \omega)$. This is done  in order
to identify  the regions where there is Landau damping. Clearly, the zeros of
\ref{EQ:DETERMINANTSINGLE} yield both damped and undamped plasmon
excitations.  In order to isolate the undamped modes, it is convenient to consider the
plasmon spectral weight $-(1/\pi) \Im m\ \Pi (q,\omega - i \gamma_q)$, where
$ \Pi (q,\omega - i \gamma_q) =  \Pi^{(0)} (q,\omega - i \gamma_q)/\epsilon(q,\omega - i \gamma_q)$.
The plasmon relaxation rate is determined  self-consistently,
starting  with the phenomenological rate whose value hereafter is chosen as
$\hbar\gamma/ \mu = 10^{-4}$. If not for the finite relaxation
rate, the plasmon spectral weight is nonzero only in the particle-hole regions\,\cite{wunsch}.
Below, we  show that this condition applies only
for the single layer configuration.
The concept is  demonstrated in \ref{FIG:1}(b.1). The naturally damped plasmon branch
in the $1_>$ triangular region becomes strongly Landau damped once it enters the
particle-hole mode  region $2_>$.

\par
By dressing Dirac electrons with  circularly polarized photons, they acquire an
effective mass and  a gap $E_g$ is opened between the valence and conduction
bands.  The gap
is determined by the intensity and frequency of the laser field. The presence
of $E_g$ in turn induces a gap in the particle-hole mode region  for
$q/k_F \stackrel{>}{\sim}1$.  The particle-hole free region $5_>$ widens with
increasing $E_g$. Its contribution to the plasmon region function also grows,
as indicated by the second term in \ref{EQ:PLASMONREGIONFUNCTION}.
Meanwhile, the Dirac  contribution of the plasmon region function is
decreased through the first term in the same equation.  Schematically,
the opening of the plasmon window by the electron dressing is shown
in \ref{FIG:6}. There is one more region in the particle-hole
continuum which gets  opened by the energy gap $E_g$,
namely region $4_<$. 
We  give it special attention since this region provides the
main difference between  dressed Dirac electrons and  conventional graphene.
In the latter case, this region simply merges with $5_>$, since the
$\omega = \hbar v_F q$ line has no physical meaning. However, in gapped graphene,
this region is  very special due to the possible natural plasmon gain existing in it.
Indeed, according to \ref{EQ:POLARIZATION}, the  polarization acquires an
imaginary part provided analytical continuation
 $\omega \rightarrow \omega - i \gamma$ from the real axis into the lower complex half-plane $\gamma >0$.
The imaginary part of the polarization becomes a weighted 
delta function, centered at the plasma frequency 
$\Im m \Pi (q,\omega) = \Im m (1/ (v_c(q) \epsilon(q,\omega_p))) = 
\text{sgn} (\Im m  \epsilon (q,\omega_p)) (\pi/v_c(q)) \delta (\Re e \epsilon (q,\omega_p))  =  \pi W(q,\omega_p) \delta(\omega - \omega_p)$.
\par
In the  region $5_>$ the plasmon weight is the conventional positive one given by
$W(q,\omega_p) = W_0 5_>$ , 
with $W_0 =  1/ v^2_c (q) \frac{\partial}{\partial \omega} \Re e  \Pi^{(0)} (q,\omega) \vert_{\omega=\omega_p} )>0$.
This behaviour can be seen on the lower insets of \ref{FIG:1} (a.3).
However, when the plasmon branch enters the region $4_>$
the plasmon weight may flip its sign
$W(q,\omega_p) = - W_0 4_>$,
since $\Im \Pi^{(0)}(q,\omega_p) <0$
as indicated on the upper inset of \ref{FIG:1}(a.3).
Interestingly enough the change in sign of the plasmon weight requires the plasmon branch to cross $\omega = \hbar v_F q$ line.
On one hand, the gapeless graphene does not show the crossover
from the plasmon loss to the gain since the above condition is never satisfied.
On the other hand, if one is to increase energy gap ($E_g/\mu \approx 2$) so that the plasmon branch 
is completely forced into the $4_<$ region, then the plasmons also show only the loss and the positive plasmon weight
$W(q,\omega_p) = W_0 4_>$.
 For even larger energy gap, the  plasmons are forced into region $2_<$  and become
fully Landau damped.
The plasmon dispersion for a single layer of gapped graphene is presented in \ref{FIG:1}(a.2), (a.3).
Numerical results shown in \ref{FIG:1}(b.2), (b.3) demonstrate the predominant plasmon
 gain  for short wavelengths.  
In the long wave length limit the plasmon branches are always well behaved.
Therefore, at this level of theory, it is impossible to say if the anomalous plasmon 
behaviors has some physical meaning or being just an indicator of RPA failure at larger $q$. 

\par
\begin{figure}
\centering
\includegraphics[width=0.7\columnwidth]{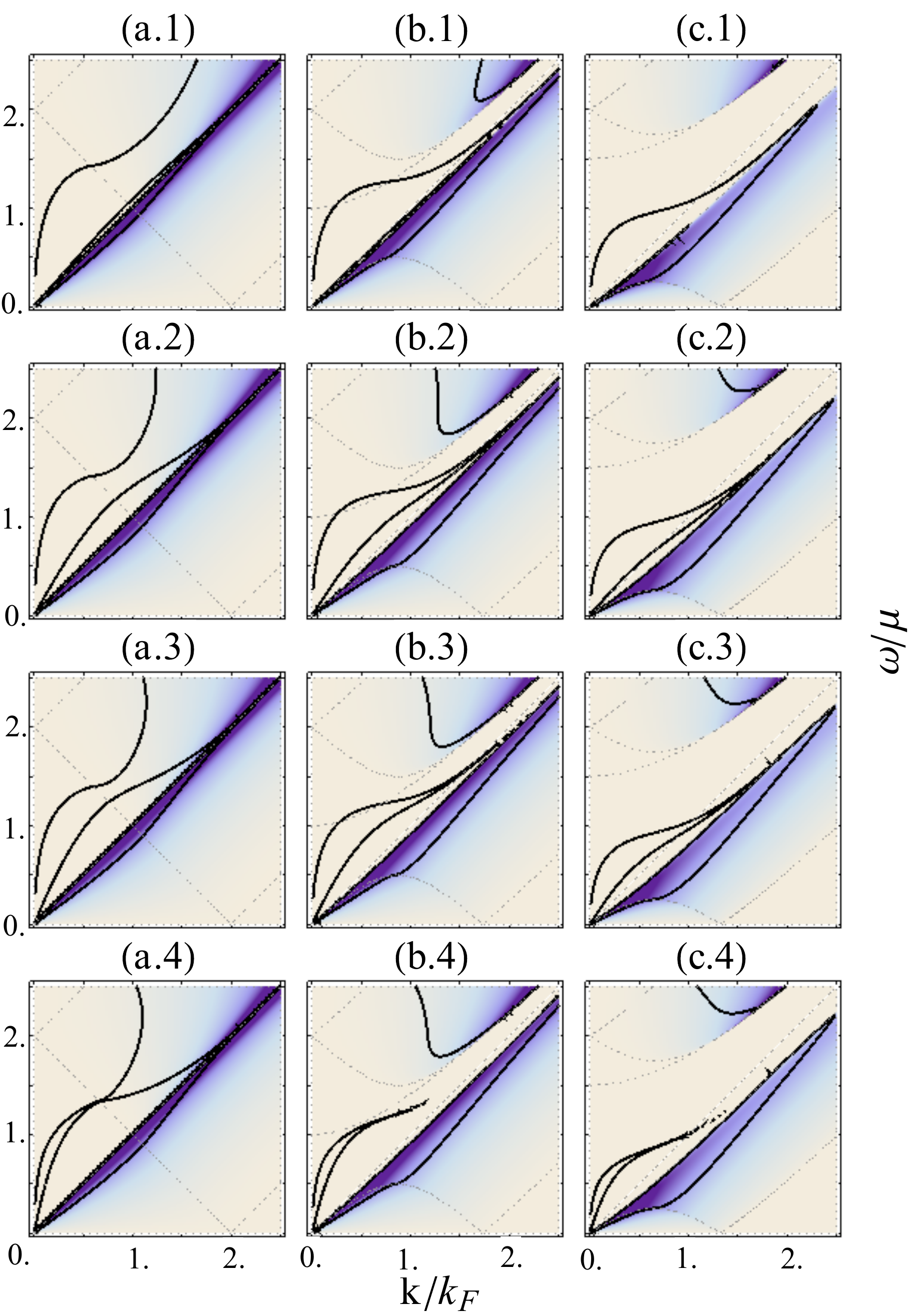}
\caption{ (Color online)  Plasma excitations and damping (analogous to  column (a) of
\ref{FIG:1}) for  double layer configuration  of graphene layers. The imaginary
part of the polarization  $\Pi^{(0)}$ is plotted as a function of frequency and wave
vector.  The solid black lines are the plasmon dispersion curves.
Columns (a,\,b,\,c)  are for $E_g/\mu = \{{0.0,\,1.0,\,1.5}\}$, respectively.
The induced gap and chemical potential are the same for both layers (symmetric case).
Rows (1,\,2,\,3,\,4) depict plasmon branches for  inter-layer distances $k_Fd = \{{0.1,\,0.5,\,1.0,\,5.0}\}$,
respectively.}
\label{FIG:2}
\end{figure}
\begin{figure}
\centering
\includegraphics[width=0.7\columnwidth]{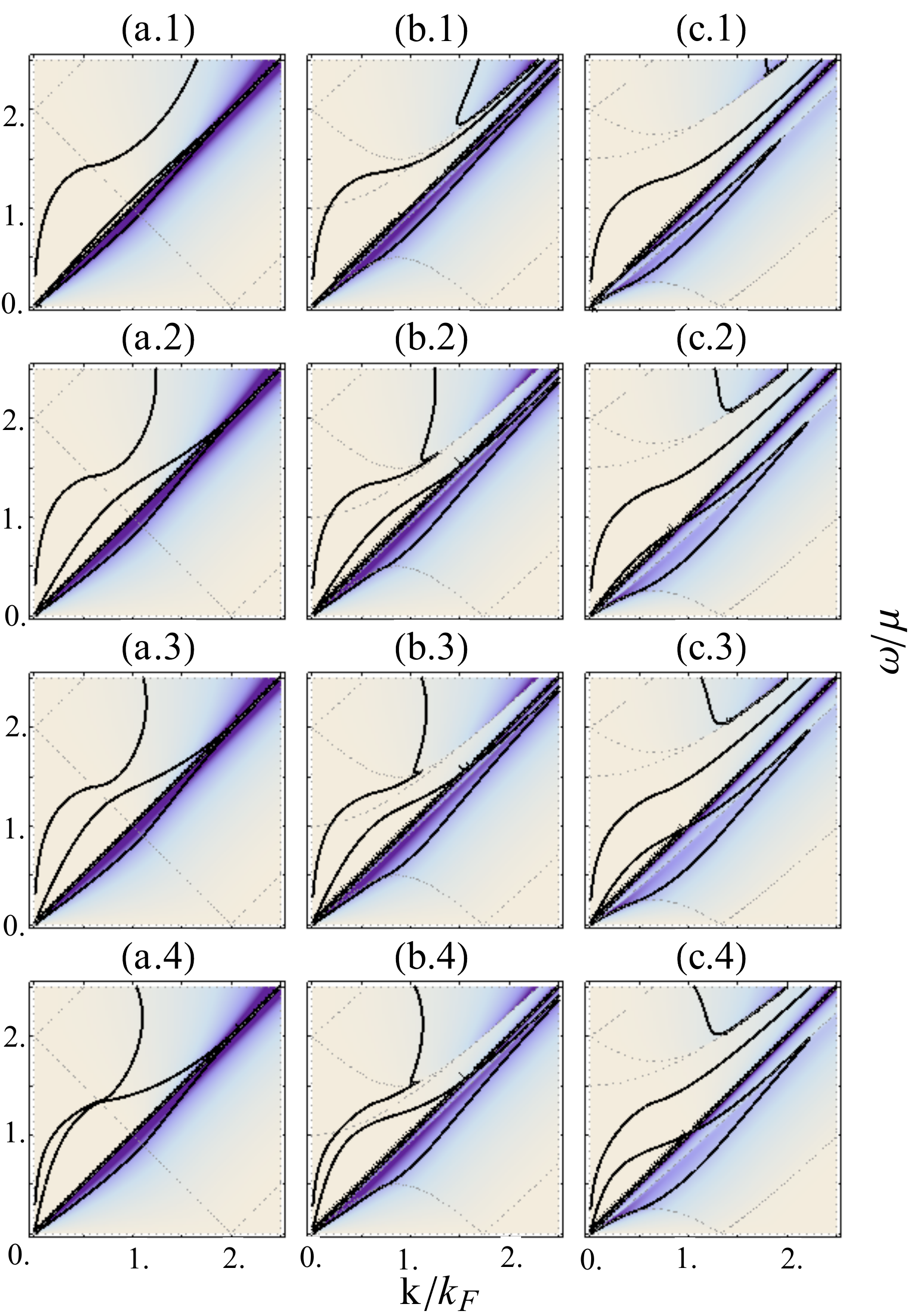}
\caption{ (Color online)  Plasma excitations and damping (analogous to  column (a) of
\ref{FIG:1} for   graphene double layers. The difference with \ref{FIG:2}, is that the energy gap $E_g$ may not be the same for both layers.
Columns  (a,\,b,\,c) correspond to an induced gap $E_g/\mu = \{{0.0,\,1.0,\,1.5}\}$, respectively, only
for one of the layers. Although the chemical potential is the same for both layers, a gap is not
induced on  the second layer (asymmetric case). Rows (1,\,2,\,3,\,4) show plasmon branches for  inter-layer
separations $k_Fd = \{{0.1,\,0.5,\,1.0,\,5.0}\}$.}
\label{FIG:3}
\end{figure}
\begin{figure}
\centering
\includegraphics[width=0.7\columnwidth]{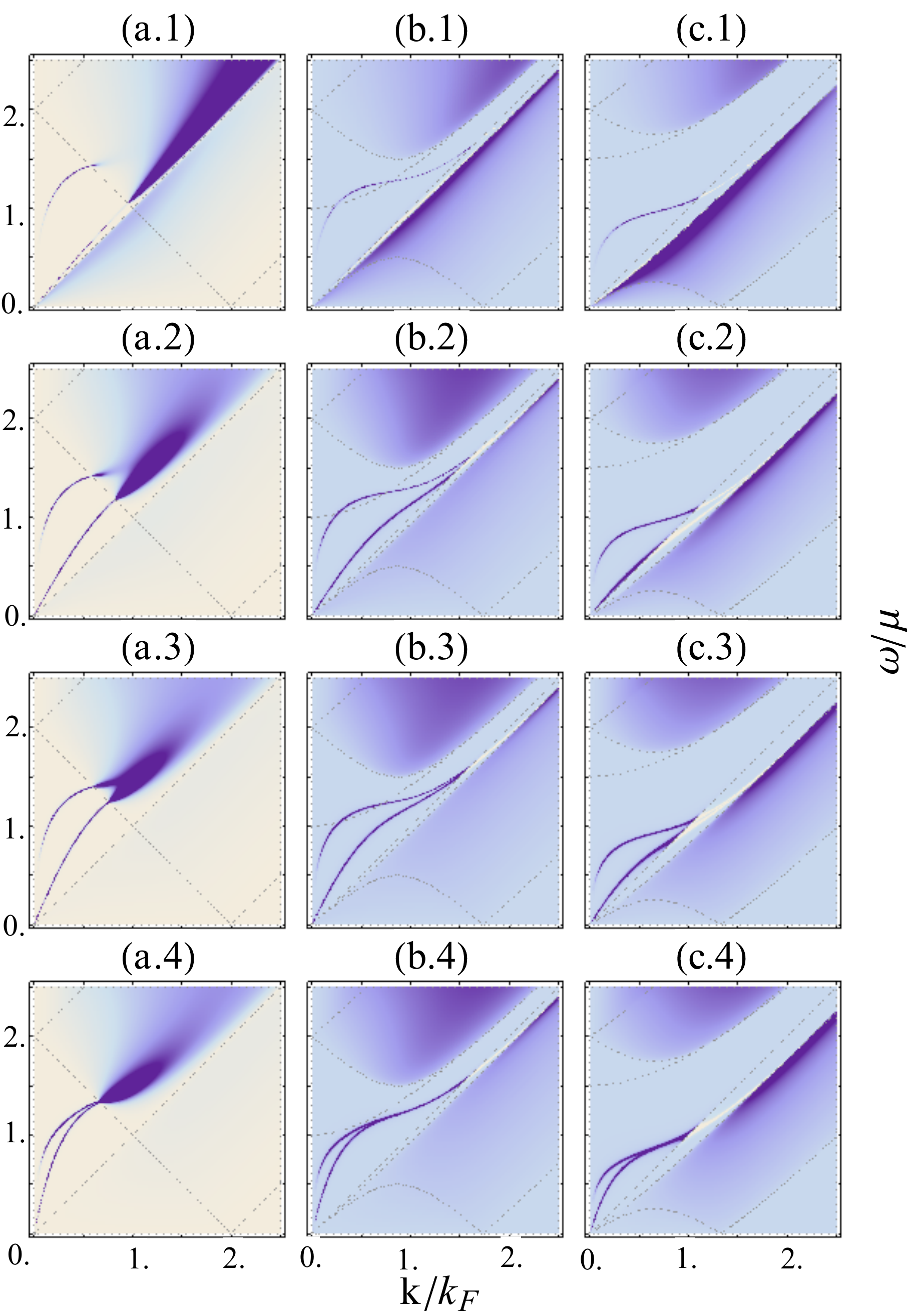}
\caption{ (Color online)
Plots similar to panels (b) in \ref{FIG:1} showing $\Im m\ \Pi^{(0)}_{11}$ for double graphene layer configuration.
For columns (a,\,b,\,c), we chose $E_g/\mu = \{{0.0,\,1.0,\,1.5}\}$, respectively.
The induced gap and chemical potential are the same for both layers (symmetric case).
Rows (1,\,2,\,3,\,4) display plasmon branches for  inter-layer distances $k_Fd = \{{0.1,\,0.5,\,1.0,\,5.0}\}$.}
\label{FIG:4}
\end{figure}
\begin{figure}
\centering
\includegraphics[width=0.7\columnwidth]{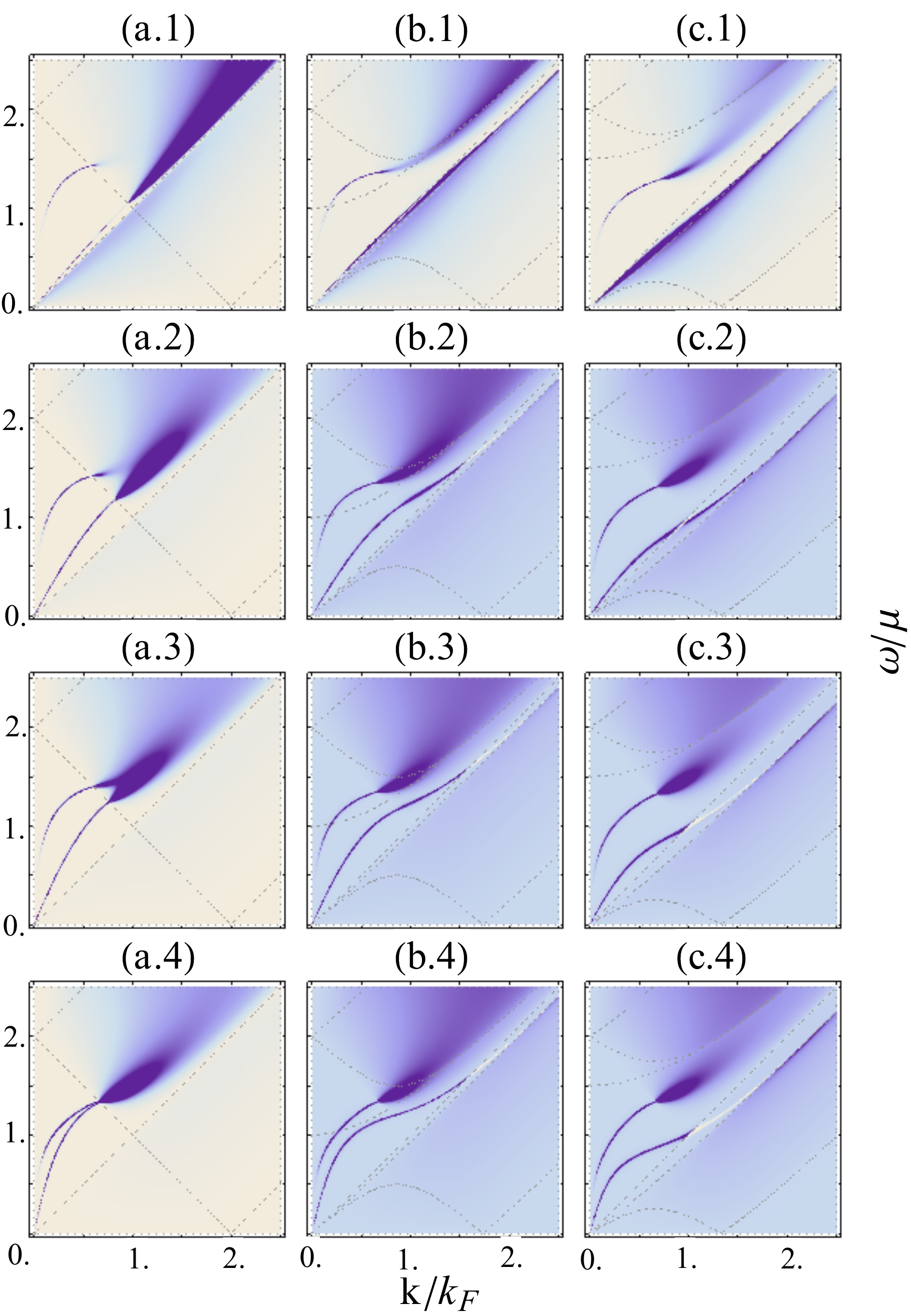}
\caption{ (Color online)
Plots similar to those in column (b) of \ref{FIG:1} showing $\Im m\ \Pi^{(0)}_{11} $
for a pair of graphene layers.
Columns (a,\,b,\,c) were obtained using  $E_g/\mu = \{{0.0,\,1.0,\,1.5}\}$, respectively.
Although the chemical potential is uniform for both layers, a  gap is not induced on
the second layer (asymmetric case).
Panels (1,\,2,\,3,\,4) show plasmon branches for inter-layer
distances $k_Fd = \{{0.1,\,0.5,\,1.0,\,5.0}\}$.}
\label{FIG:5}
\end{figure}

Turning now to the double layer configuration, we present in \ref{FIG:2} 
through \ref{FIG:5} the plasmon dispersion and damping
for the symmetric and asymmetric cases. In these plots, the panels
(a.1, a.2, a.3, a.4)  correspond to zero energy gap for chosen inter-layer
separation. In the presence of an energy gap,  we note that
one of the plasmon modes (the acoustic
branch $\omega_-$) may be forced into region $4_<$  for values of
$E_g/\mu$ as small as one (See \ref{FIG:2,FIG:3}(b.1, c.1)).
As far as the symmetric configuration is concerned,  this mode demonstrates
\emph{gain}, while the asymmetric case shows losses in that region as may be deduced
from its spectral weight in \ref{FIG:4,FIG:5} (b.1, c.1).

\par
For the symmetric double layer, there are two Landau damped plasmon modes
in the region $2_<$.
One of them runs just along $2_<,4_<$ interface.
While the
other starts along $2_<,1_<$ interface and then eventually merges with the first branch at short wavelength,
as demonstrated in \ref{FIG:2}.
The frequencies of these  plasmon branches are almost
independent of  the inter-layer separation $d$. The asymmetric configuration in
effect causes one of these damped modes to become purely acoustic with frequency
$\omega \sim q v_F$  over a wide range of wave vector $q$,  as shown in \ref{FIG:3}.

\par
Let us now confine our attention to  the plasmon branches whose frequencies
in the long wavelength limit coincide with those of the bonding  ($\omega_+$)
and antibonding ($\omega_-$)  modes.  For large inter-layer distance
satisfying $k_F d \gg 1$,
in the absence of an energy gap, $E_g= 0$,  these modes  converge onto
the  line   $\omega + v_F q = 2 \mu/\hbar$. By increasing the gap for the symmetric
double layer, there emerge    Landau damped plasmon branches in regions $2_>, 4_>$
from the bonding mode, as shown in \ref{FIG:2}(b,c).
The remaining part of the $\omega_+$ branch is almost  independent of $d$.
This part of the excitation branch  mimics the single layer plasmon subjected to $E_g$
(symmetric configuration, \ref{FIG:2}(b.4, c.4)) or gap independent (asymmetric
case, \ref{FIG:3} (b.4,c.4)). The latter case is of special interest since it obeys
the conventional plasmon dispersion law for graphene while seemingly progressing
into the gap within the particle-hole mode region. However, its spectral weight
as seen in \ref{FIG:5}  demonstrates strong Landau-like damping
once it enters region $\omega + v_F q > 2 \mu/\hbar$.

\par
By  bringing the layers close to each other, $\omega_-$ becomes  acoustic like
(see \ref{EQ:DLACOUSTICAL}), and ,for small
inter-layer distance, may be mostly accommodated by region $4_<$.
The $\omega_+$ mode is similar to  that for  the single layer.
That is, $\omega_+  \approx \omega_p (E_g )$ for the symmetrical case  \ref{EQ:DLACOUSTICAL} ,
and is gap independent in the asymmetrical case with $\omega_+  \approx \omega_p (E_g=0)$.

\section{IV. Concluding Remarks}
\label{sec4}

Theoretical results reported here were carefully presented to stimulate
experimental verification for both the single and double  layer configuration.
One possible probe which may be employed is EELS using an electron
spectrometer\,\cite{Ibach}. In EELS, some of the electrons undergo inelastic scattering,
losing energy and having their paths slightly modified.  The inelastic interactions
include phonon excitations,  inter band and intra band particle-hole excitations,
plasmon excitations and Cerenkov radiation\,\cite{antonio}.  A separate study is needed to
calculate the stopping power due to these mechanisms in order to supplement
the results reported in this paper.

\acknowledgement

This research was supported by  contract \# FA 9453-07-C-0207 of AFRL. DH would like
 to thank the Air Force Office of Scientific Research
(AFOSR) for its support. 
We also appreciate useful discussions with P. Pytkovskiy.

\newpage

\newpage
\begin{thebibliography}{23}
\bibitem{kibis}O. V. Kibis, Phys. Rev. B \textbf{81}, 165433 (2010).

\bibitem{SOI} X.-F. Wang and T. Chakraborty, Phys. Rev. B. \textbf{75}, 033408 (2007).

\bibitem{SOI-2}P. K. Pyatkovskiy,  J. Phys.: Condens. Matter \textbf{21},  025506 (2009).

\bibitem{li_2009}G. Li, A. Luican, and E.Y. Andrei, Phys. Rev. Lett. {\bf 102}, 176804 (2009).

\bibitem{giovannetti_2007}G. Giovannetti, et. al.,  Phys. Rev. B. {\bf 76}, 073103 (2007).

\bibitem{wunsch}B. Wunsch, T. Stauber, F. Sols, and F. Guinea,  New J. Phys. \textbf{8}, 318 (2006).

\bibitem{shung1}K. W-K. Shung,  Phys. Rev. B. {\bf 34}, 979 (1986).

\bibitem{shung2}K. W-K. Shung, Phys. Rev. B. {\bf 34}, 1264 (1986).

\bibitem{Lin}J.-Y. Wu, S.-C. Chen, O. Roslyak, G. Gumbs, and M.-F. Lin, ACS Nano (Web), DOI: 10.1021/nn1024847 (2011).

\bibitem{gumbs1}G. Gumbs,  Phys. Rev. B.  {\bf 70}, 235314 (2004).

\bibitem{manvir}M. S. Kushwaha,  Phys. Rev. B. {\bf 76}, 245315 (2007).

\bibitem{gumbs2}G. Gumbs,  Phys. Rev. B. {\bf 73}, 165315 (2006).

\bibitem{rotenberg_2008}E. Rotenberg, et. al., Nature Matter. {\bf 7}, 258 (2008).

\bibitem{kim_2008}S. Kim, J. Ihm, H. J. Choi, and Y. Son, Phys. Rev. Lett. {\bf 100}, 176802 (2008).

\bibitem{zhou_2007}S. Y. Zhou et al., Nature Matter. {\bf 6}, 770 (2007).

\bibitem{bostwik_2007} A. Bostwick et al., Nature Phys. {\bf 3}, 36 (2007).

\bibitem{p6.1}B. N. J. Persson, Solid State Commun.  \textbf{52}, 811 (1984).

\bibitem{ggumbs}G. Gumbs,  Phys. Rev. B.  {\bf 37}, 10184 (1988).

\bibitem{antonio}A. Balassis and G. Gumbs, J. Appl. Phys. \textbf{106}, 103102 (2009).

\bibitem{EELS_SSC}G. Gumbs, Solid State Commun. {\bf 65}, 393 (1988).

\bibitem{quaimzadeh_2009} A. Quaimzadeh and R. Asgari, New J. Phys. {\bf 11}, 095023 (2009).

\bibitem{dassarma}S. Das Sarma and A. Madhukar,  Phys. Rev. B. \textbf{23}, 805 (1981).

\bibitem{Ibach}H. Ibach and D. L. Mills, {\em Electron Energy loss spectroscopy and
surface vibrations\/}, Academic Press, New York, (1982).
\end{thebibliography}
\end{document}